\documentclass[aps,prl,twocolumn,showpacs,preprintnumbers,amsmath,amssymb, superscriptaddress]{revtex4}

\usepackage{graphicx}
\usepackage{dcolumn}
\usepackage{bm}

\begin{document} 

\title{An Entanglement Filter}


\author{Ryo Okamoto}
\altaffiliation[]{These authors contributed equally}
\affiliation{Research Institute for Electronic Science, Hokkaido University, Sapporo 060-0812, Japan}
\affiliation{The Institute of Scientific and Industrial Research, Osaka University, Mihogaoka 8-1, Ibaraki, Osaka 567-0047, Japan}
\author{Jeremy L. O'Brien}
\altaffiliation[]{These authors contributed equally}
\affiliation{Centre for Quantum Photonics, H. H. Wills Physics Laboratory \& Department of Electrical and Electronic Engineering, University of Bristol, Merchant Venturers Building, Woodland Road, Bristol, BS8 1UB, UK}
\author{Holger F. Hofmann}
\affiliation{Graduate School of Advanced Sciences of Matter, Hiroshima University, Hiroshima 739-8530, Japan}
\author{Tomohisa Nagata}
\affiliation{Research Institute for Electronic Science, Hokkaido University, Sapporo 060-0812, Japan}
\affiliation{The Institute of Scientific and Industrial Research, Osaka University, Mihogaoka 8-1, Ibaraki, Osaka 567-0047, Japan}
\author{Keiji Sasaki}
\affiliation{Research Institute for Electronic Science, Hokkaido University, Sapporo 060-0812, Japan}
\author{Shigeki Takeuchi}
\altaffiliation[Electronic address: ]{takeuchi@es.hokudai.ac.jp}
\affiliation{Research Institute for Electronic Science, Hokkaido University, Sapporo 060-0812, Japan}
\affiliation{The Institute of Scientific and Industrial Research, Osaka University, Mihogaoka 8-1, Ibaraki, Osaka 567-0047, Japan}

\begin{abstract}
The ability to filter quantum states is a key capability in quantum information science and technology, where one-qubit filters, or polarizers, have found wide application. Filtering on the basis of entanglement requires extension to multi-qubit filters with qubit-qubit interactions. We demonstrate an optical entanglement filter that passes a pair of photons if they have the desired correlations of their polarization. Such devices have many important applications to quantum technologies. 
\end{abstract}

\maketitle


Filters, that pass the desired and reject the unwanted (material, signal, frequency, polarization, etc) are one of the most important scientific and technological tools available to us. Quantum information science and technology is concerned with harnessing quantum mechanical effects to gain exponential improvement and new functionality for particular tasks in communication \cite{gi-rmp-74-145}, computation \cite{nielsen}, measurement \cite{gi-sci-306-1330}, and lithography \cite{bo-prl-85-2733}. Perhaps the most unique of these quantum mechanical features is entanglement. Filters which act on the quantum correlations associated with entanglement must operate non-locally on multiple quantum systems, typically two-level qubits. Such a device has been proposed for photonic qubits \cite{ho-prl-88-147901}, however, the technical requirements to build such a device, an optical circuit with two ancillary photons and multiple quantum gates, requiring both quantum interference and classical interference in several nested interferomters, have been lacking.

We demonstrate an entanglement filter by combining two key recent technological approaches---a displaced-Sagnac architecture \cite{na-sci-316-726} and partially polarizing beam splitters \cite{la-prl-95-210504, ki-prl-95-210505, ok-prl-95-210506}. The entangling capability of the filter was verified, distinguishing it from classical ones. As our entanglement filter acts on photonic qubits, it is promising for quantum technologies, as photons are the logical choice for communication \cite{gi-rmp-74-145,gi-nphot-1-165}, metrology \cite{na-sci-316-726,hi-nat-450-393} and lithography \cite{da-prl-87-013602}, and are a leading approach to information processing \cite{ob-sci-318-1567}. The filter can be used for the creation as well as the purification of entanglement \cite{pa-nat-410-1067,ya-nat-421-343}, which will be important in realizing quantum relays and repeaters for long distance quantum communication. 

The most common examples of a filter for photonic qubits is a polarizer that transmits only the horizontal component of the incident photons (Fig. 1A). Polarization filters are crucial for the initialization and read out of photonic qubits as well as for the measurement related effects such as the quantum Zeno effect \cite{mi-jmp-18-756, kw-prl-83-4725}. More sophisticated photonic filters that transmit only single photon states while rejecting higher photon number states have also been realized by using ancillary photons and photon detection to induce the required optical non-linearity \cite{sa-pra-71-021801,sa-prl-96-083601,re-prl-98-203602}. The concept of a polarization filter can be extended to a higher dimensional Hilbert space of multi-qubits and one may anticipate that the necessary qubit-qubit interactions required will also require the use of ancillary photons.

The two-photon polarization filter proposed in \cite{ho-prl-88-147901} is a device that transmits photon pairs only if they share the same horizontal or vertical polarization without decreasing the quantum coherence between these two possibilities (Fig. 1B). As this filter is not sensitive to the individual polarization of the photons, it can preserve and even create entanglement between the photons passed through it (Fig. 1C). The filter achieves this two qubit filtering effect by using two ancilla photons as probes that detect whether the two input photons are in the desired states or not. Detection of the ancilla photons indicates a successful filter operation, requiring no post-selection in the output. Note that this is a significant difference between this filter and the function realized by output post-selection on a polarizing beam splitter \cite{pa-nat-410-1067, pi-pra-64-062311}, where it is necessary to eliminate the cases where both photons exit on the same side. Because our filter should reliably produce the correct two photon output in separate ports of the device, it can be integrated into larger quantum information networks, just like a one-qubit polarization filter. It is therefore possible to apply it to a wide range of problems, such as encoding quantum information in multiple-qubits and entanglement purification from mixed entangled states.

\begin{figure}
\center{\includegraphics[width=1\linewidth]{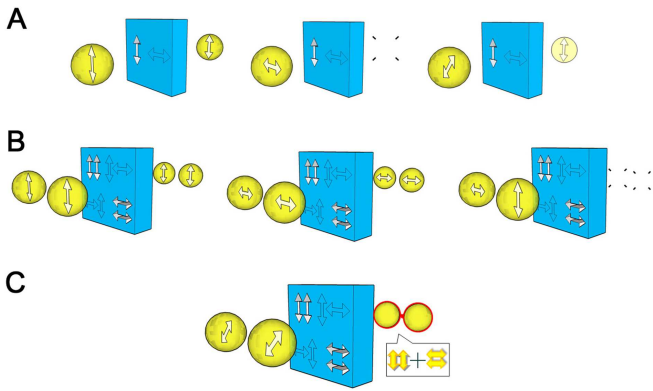}}
\caption{The function of a polarization filter and a entanglement filter. (A) Polarization filter pass only the certain polarization component of single photons. (B) Entanglement filter pass a pair of photons only if they share the same horizontal or vertical polarization. (C) As the quantum coherence between these two possibilities is preserved during the process, the output state is entangled when two diagonally polarized photons are input.}
\end{figure} 

%

The operation of this filter can be summarized by a single operator $\hat{S}$ describing the effect on an arbitrary two photon input state \cite{ho-prl-88-147901}:
\begin{equation}
\hat{S}=\frac{1}{4}({|HH\rangle \langle HH|-|VV\rangle \langle VV|}),
\end{equation}
where $|HH\rangle$ and $|VV\rangle$ denotes the state where both incident photons are horizontally ($H$) and vertically ($V$) polarized, respectively. This operator filters out the $|HV\rangle$ and $|VH\rangle$ components of the two photon state without reducing the coherence between the remaining $|HH\rangle$ and$|VV\rangle$ components. The negative sign between the terms is a consequence of the phase difference between two photon reflection and two photon transmission at the beam splitters. The factor of 1/4 is an expression of the transmission
probability of 1/16 for $|HH\rangle$ and $|VV\rangle$. It should be noted however that the detection of the ancilla photons indicates a successful transmission of the two input photons, so that it is not necessary to detect the output photons in order to know that they were successfully transmitted.

To understand the unique quantum properties of this filter, it is useful to consider the effects of the filtering process on two diagonally polarized photons. As diagonal polarizations are coherent superpositions of horizontal and vertical polarizations ($|P\rangle\equiv\frac{1}{\sqrt{2}}(|H\rangle+|V\rangle)$; $|M\rangle\equiv\frac{1}{\sqrt{2}}(|H\rangle-|V\rangle)$), two diagonally polarized photons have a well defined coherence between their $|HH\rangle$ and $|VV\rangle$ components that determines whether their diagonal polarizations are parallel ($+$) [\emph{eg}: $|PP\rangle ~/~ |MM\rangle = \frac{1}{2}(|HH\rangle+|VV\rangle\pm ...)$] or orthogonal ($-$) [\emph{eg}: $|PM\rangle ~/~ |MP\rangle = \frac{1}{2}(|HH\rangle-|VV\rangle \pm ...)$]. The operator $\hat{S}$ preserves the magnitude of this coherence, but flips the sign: Eq. 1. Therefore, the correlation between the diagonal polarizations is inverted; parallel inputs resulting in the superpositions of the orthogonal outputs and vice versa. Moreover, the coherence between $|HH\rangle$ and $|VV\rangle$ indicates a correlation between the circular polarizations---opposite directions ($+$) or same direction ($-$). Thus, filtering out the $|HV\rangle$ and $|VH\rangle$ components of two diagonally polarized input photons also generates correlations between the circular polarizations of the photons, indicating the generation of entanglement.

A schematic of the optical quantum circuit for the entanglement filter Fig. 2A shows that multi photon interferences occur at each of the four beam splitters BS1 to BS4. 
The crucial ones are BS2 and BS3. The working principle of the quantum circuit is as follows: whenever one of the ancilla photons meets a single input photon at BS2 or BS3, two photon quantum interference between the reflection of both photons and the exchange of ancilla and input photon reduces the probability of finding a single photon at D1 or D2 to zero (Fig. 2A, inset) \cite{ho-prl-59-2044}. Thus, a single horizontally polarized photon cannot pass the circuit, resulting in the elimination of the $|HV\rangle$ and $|VH\rangle$ components. However, the $|HH\rangle$ component can pass because the two input photons will both arrive at BS2 or BS3
together (the two H photons undergo quantum interference at BS1). In that case, the negative probability amplitude corresponding to the exchange of the ancilla photon and one of the input photons is higher than the positive probability amplitude for three photon reflection, and the $|HH\rangle$ component is transmitted with a flipped phase. 

\begin{figure*}
\center{\includegraphics[width=1\linewidth]{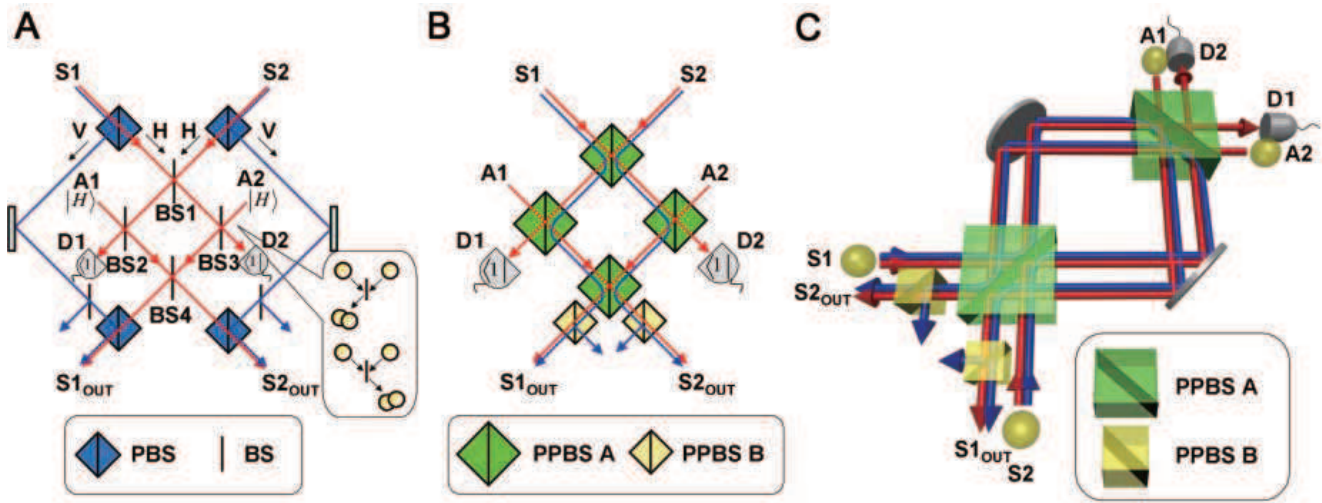}}
\caption{Optical quantum circuit for the nondestructive entanglement filter. (A) The original circuit includes concatenated path-interferometers together with four quantum interferences. PBS: Polarizing Beam Splitter which reflects (transmits) vertical (horizontal) polarization component, BS: Beam Splitter whose reflectance/transmittance is 1/2 for horizontal/vertical polarization. (inset) two photon interference at BS2 and BS3. Blue (red) line indicates optical paths for vertically (horizontally) polarized components. (B) A semi-simplified circuit using partially polarizing beam splitters (PPBS). The reflectance of PPBS A for horizontally and vertically polarized photons are 1/2 and 1, respectively. The transmittance of PPBS B for horizontally and vertically polarized photons are 1/2 and 1, respectively. (C) The final simplified system used in our experiment. Now the system includes one super-stable optical path-interferometer realized using displaced Sagnac architecture.}
\end{figure*} 


To realize the quantum optical circuit of Fig. 2A, we have combined two recent technological approaches in order to simplify and stabilize quantum optical circuits: the displaced-Sagnac architecture \cite{na-sci-316-726} and partially polarizing beam splitters (PPBSs) \cite{la-prl-95-210504, ki-prl-95-210505, ok-prl-95-210506}. The PPBSs reflect vertically polarized photons perfectly while transmitting/reflecting horizontally polarized photons with 50\% probability. Replacing all the beam splitters (BS1 to BS4) with such PPBSs, we can remove the four polarizing beamsplitters used to separate the two polarizations (Fig. 2B). The remaining optical path interferometer is realized as a displaced Sagnac interferometer (Fig. 2C). In this setup, all the four polarization modes of two input photons pass through all the optical components inside the interferometer so that the path-differences between those four polarization modes are robust against drifts or vibrations of optical components.

We used photons generated via type-I spontaneous parametric down-conversion \cite{bu-prl-25-84}. The pump laser pulses (82 MHz, @ 390 nm)  pass through a beta-barium borate (BBO) crystal twice to generate two pairs of photons. Two photons in one of the pairs are used as the signal photons, and the two photons in the other pair are used as the ancillary photons. The visibility of the Hong-Ou-Mandel dip \cite{ho-prl-59-2044} was $96\pm1 \%$ for two photons in the same pair, and $85\pm5 \%$ for photons from different pairs. To test the performance of our quantum filter circuit, we used coincidence measurements between the four threshold detectors at D1, D2 and the two outputs, rather than using photon number discriminating detectors \cite{ta-apl-74-1063,ki-apl-74-902,ga-nphot-1-585,ka-nphot-2-425} for D1 and D2, as we needed to analyze the polarization state of the output to confirm correct operation. This polarization analysis was performed using pairs of half wave plates and quarter wave plates together with polarizing beam splitters.

%
First, we checked the essential operation of the filter circuit: we prepared input signal photons in the four combinations of horizontal (H) and vertical (V) polarizations (which we call the Z-basis states) and observed how those input states are filtered by the circuit (Fig. 3). It is clear from the experimental data that the photon pairs are transmitted through the filter when the two input photons share the same polarization (HH or VV), and most of the pairs are filtered out when the two input photons have different polarization (HV or VH). The fidelity of this process can be defined as the ratio of correctly transmitted photon pairs to the total number of transmitted photon pairs. For the filter's operation on horizontal and vertical polarizations, the fidelity is $F_{Z\rightarrow Z}=0.80$. 

Next, it is necessary to test the preservation of quantum coherence between the $|HH\rangle$ and the $|VV\rangle$ components transmitted by the filter. As explained above, this can be done by using diagonally polarized photons in the input. We prepared pairs of photons in all combinations of diagonal linear polarizations $|P\rangle$ and $|M\rangle$ (the X-basis states) as signal inputs. The input photons are then in an equal superposition of the four different horizontal/vertical polarization states $|HH\rangle$, $|HV\rangle$, $|VH\rangle$, and $|VV\rangle$ (with different signs of the coherences for each of the four different inputs). The filter should transmit only the HH and the VV components while preserving the quantum coherence between them. As a result, the ideal output states are entangled states with maximal correlations in both the circular and the diagonal polarizations. 
To test this entanglement generation, we first detected the output photons in the right-circular $|R\rangle\equiv (|H\rangle+i|V\rangle)/\sqrt{2}$ and left-circular $|L\rangle\equiv (|H\rangle-i|V\rangle)/\sqrt{2}$ polarizations  (Y-basis). The predicted correlations between the circular polarizations of the output are observed with a fidelity of $F_{X\rightarrow Y}=0.68$. To complete the test of entanglement generation, we also detected the diagonal polarizations of the output to test whether the polarization correlation is correctly flipped by the filter. The fidelity of this filter operation is  $F_{X\rightarrow X} =0.60$. Taken together, the three experimental tests are sufficient for the evaluation of the quantum filter operation.  Because almost all the errors conserve horizontal/vertical polarization (see Appendix 1), the probability of the correct operation $\hat{S}$ is given by the process fidelity 
\begin{equation}
F_p = (F_{z->z}+F_{x->y}+F_{x->x}-1)/2 = 0.54.
\end{equation}
The entanglement capability of the filter can be evaluated as $C=2 F_p-1 = 0.08$, and so our experimental results provide clear evidence of the entangling capability of the quantum filter operation.

\begin{figure}
\center{\includegraphics[width=1\linewidth]{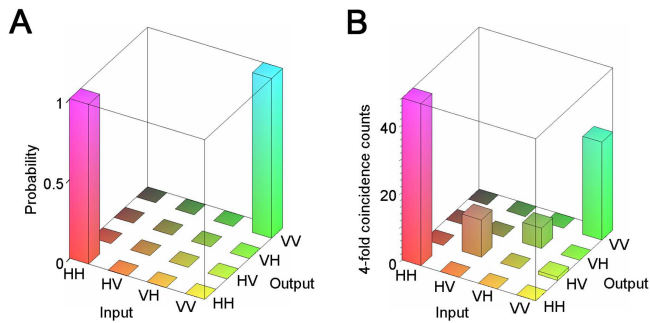}}
\caption{Experimental results. Input signal photons are prepared in horizontal (H) or vertical (V) polarization and are measured on H/V basis. (A)Theoretically predicted probabilities (B) four fold coincidence count rates [counts/800s] of the four detectors D1 to D4 are shown. Note that the events where two pairs of photons are simultaneously incident to the ancillary inputs and no photons are incident to the signal inputs are subtracted, as confirmed by a reference experiment without input photons. ( 6 coincidence counts / 800s for each of four cases with HH outputs. ) }
\end{figure}

The entanglement filter will be a key element in the control of multi-photon quantum states, with
a wide range of applications in entanglement based quantum communication and quantum information processing. For the present tests of the performance of the quantum filter circuit, we used threshold detectors to monitor the output state. For applications where the output state cannot be monitored, high-efficiency number-resolving photon detectors \cite{ta-apl-74-1063,ki-apl-74-902,ga-nphot-1-585,ka-nphot-2-425} could be used at D1 and D2 to generate the heralding signals. Such a circuit could be used for on-demand generation of entangled photons, non-destructive entanglement purification, and could be implemented using an integrated waveguide architecture \cite{po-sci-320-646}. This work was supported by the Japan Science and Technology Agency (JST), Ministry of Internal Affairs and Communication (MIC), Japan Society for the Promotion of Science (JSPS), 21st Century COE Program, Special Coordination Funds for Promoting Science and Technology.
\\

\clearpage

\section*{Appendix 1}
\textit{Derivation of the process fidelity.} 
The errors due to the change of input horizontal/vertical polarization
(i.e. $|HH\rangle \to |HV\rangle$) are negligibly small in our experiment shown in
Fig. 3B ( $Z \to Z$ measurement ). Based on this fact, we can assume that
the input horizontal/vertical polarization is preserved in our filter,
and thus our filter operation can be well described in terms of
superpositions of $|HH \rangle \langle HH|$, $|HV  \rangle \langle HV|$, $|VH \rangle \langle VH|$ and $|VV \rangle \langle VV|$. Under this assumption, we can estimate the process fidelity of our entanglement
filter as follows.

It is convenient to express the 
errors in terms of the superpositions
\begin{eqnarray}
\hat{S}_{0}=\sqrt{2}(|HH \rangle \langle HH|-|VV \rangle \langle VV|),
\nonumber \\
\hat{S}_{zz}   =\sqrt{2}(|HH \rangle \langle HH|+|VV \rangle \langle VV|),
\nonumber \\
\hat{S}_{xy}   =\sqrt{2}(|HV \rangle \langle HV|+|VH \rangle \langle VH|),
\nonumber \\
\hat{S}_{xx}   =\sqrt{2}(|HV \rangle \langle HV|-|VH \rangle \langle VH|).
\end{eqnarray}
Here, $\hat{S}_{0}$ represents the intended operation of the entanglement filter,
$\hat{S}_{zz}$ represents the operation with a phase flip error between $|HH \rangle$ and $|VV \rangle$,
and $\hat{S}_{xy}$, $\hat{S}_{xx}$ represent the leakage (or transmission error) of the
$|HV \rangle$ and $|VH \rangle$ components, either with or without a phase flip between these components.
The operation of our filter can then be written as 
\begin{equation}
\label{eq:pm}
E(\rho_{in})= \sum_{n,m} \chi_{nm} \hat{S}_n \rho_{in} \hat{S}_m
\end{equation}
where $n,m \in \{0, zz, xy, xx\}$ and $\chi_{nm}$ are the process matrix elements
of the noisy quantum process.

Each of our experimentally observed truth table operations $i\to j$ is 
correctly performed by
$\hat{S}_{0}$ and one other operation $\hat{S}_n$, as indicated by the index $n=ij$ for $i,j \in \{x,y,z\}$.
For example, in case of the $z\to z$ truth table, 
operation $\hat{S}_{0}$ and $\hat{S}_{zz}$
result in the correct operation and $\hat{S}_{xy}$ and $\hat{S}_{xx}$ give false results.
Therefore, the fidelities $F_{i\to j}$ can be given by the sums of the probability
$F_p=\chi_{0,0}$  for the correct operation $\hat{S}_{0}$ and the probabilities
$\eta_{ij} = \chi_{nn}$ for the errors $\hat{S}_n$ as follows.
\begin{eqnarray}
F_{z \to z}=F_p+\eta_{zz}
\nonumber
\\
F_{x \to y}=F_p+\eta_{xy}
\nonumber
\\
F_{x \to x}=F_p+\eta_{xx}
\label{eq:fid}
\end{eqnarray}
Note that these relations between the diagonal elements of the process 
matrix and the experimentally observed fidelities can also be derived from
eq. (\ref{eq:pm}) using the formal definition of the experimental fidelities. 
In this case the fidelities are determined by the sums over the correct outcomes
$|(j)_l \rangle$ in $E(|(i)_k\rangle \langle(i)_k|)$, averaged over all inputs $|(i)_k \rangle$,
\begin{eqnarray}
F_{i\to j} &=& \sum_{l,k} \langle(j)_l| E( |(i)_k\rangle \langle(i)_k| ) |(j)_l \rangle/4) \\
&=& \sum_{n,m} \chi_{nm} (\sum_{l,k} \langle(j)_l|\hat{S}_n|(i)_k \rangle \langle(i)_k|\hat{S}_m|(j)_l \rangle/4).
\end{eqnarray}
Here $k,l \in \{1,2,3,4\}$, and $(i)_k$ denotes the $k$ th state of the $i$ basis states. For example, $|(i)_1 \rangle=|HH \rangle, |(i)_2 \rangle=|HV \rangle, |(i)_3 \rangle=|VH \rangle, |(i)_4 \rangle=|VV \rangle$ for $i=z$.
The sums over initial states $k$
and final states $l$ are one for $n=m=0$ and for $n=m=ij$ and zero in 
all other cases, confirming the results in eq.(\ref{eq:fid}).

Since the diagonal elements of the process matrix correspond to the
probabilities of the orthogonal basis operations, their sum is normalized to one,
so that $\sum_n \chi_{nn}=F_p+\eta_{zz}+\eta_{xy}+\eta_{xx}=1$. It follows that the sum of all three experimentally determined fidelities is
$F_{z \to z}+F_{z \to y} + F_{x \to x} = 2 F_p+1$. Therefore, the process 
fidelity of our circuit is given by
\begin{equation}
F_p=(F_{z \to z}+F_{x \to y}+F_{x \to x}-1)/2 = 0.54.
\end{equation}

\end{document}